# Low temperature jet spectra of (DFE)₂, DFE-He, DFE-He₂ and DFE in the 2210 – 3105 cm⁻¹ region (DFE = 1,1 difluoroethylene)

A. J. Barclay,[1] A. R. W. McKellar,[2] A. Pietropolli Charmet,[3] and N. Moazzen-Ahmadi[1,*]

[1] *Department of Physics and Astronomy, University of Calgary, 2500 University Drive North West, Calgary, Alberta T2N 1N4, Canada*

[2] *National Research Council of Canada, Ottawa, Ontario K1A 0R6, Canada*

[3] *Dipartimento di Scienze Molecolari e Nanosistemi, Università Ca' Foscari Venezia, Via Torino 155, I-30172, Mestre, Venezia, Italy*

## Abstract

A pulsed supersonic slit jet expansion of a dilute mixture of 1,1 difluoroethylene (DFE) in helium is probed using a tunable infrared source to obtain spectra of (DFE)₂, DFE-He, and DFE-He₂. The DFE dimer is found to have a slipped antiparallel structure with two-fold rotational symmetry and little or no dipole moment (explaining why no pure rotational spectrum has been observed). An experimental value of 3.44 Å is determined for the separation of the monomer centers of mass. The spectra of DFE-He show line splittings due to tunneling of the He atom from one side of the DFE plane to the other. Rotational analysis of the DFE-He tunneling components in terms of a conventional asymmetric rotor yields small systematic errors due to the presence of large amplitude motions. A relatively weak spectrum is analyzed for one band of DFE-He₂, whose structure places the two He atoms in equivalent positions on each side of DFE, very close the location of He in DFE-He. Extensive spectra of DFE monomer is also obtained while searching for the cluster bands. A total of 23 bands from 2210 to 3105 cm⁻¹ are observed. These are the first such high resolution results in this region, and they are of special interest because DFE has been a prototype for the study of vibrational anharmonicity and resonances.

*Corresponding author, Email address: nmoazzen@ucalgary.ca

## I. Introduction

1,1-difluoroethylene (C$_2$H$_2$F$_2$, vinylidene fluoride, Freon1132A, 1,1-difluoroethene) is widely used as a chemical intermediate, for example in the production of fluoropolymers. It has also been a prototype for the study of vibrational anharmonicity and resonances, being large enough to be varied while still small enough to be amenable to current high level theories [1]. The results reported here were initially obtained while searching for the weakly-bound complex of 1,1-difluoroethylene (DFE) with N$_2$O in the region of the N$_2$O $\nu_1$ fundamental (2224 cm$^{-1}$), inspired by the work of Anderton et al. on the analogous DFE-CO$_2$ complex [2] to support the design of novel materials for CO$_2$ capture and storage. In addition to the desired DFE-N$_2$O spectra (which will be reported separately), we observed a number of other bands due to species which did not contain N$_2$O, namely the dimer (DFE)$_2$, the helium complexes DFE-He and DFE-He$_2$, and a large number DFE monomer bands. This represents the first spectroscopic observation of (DFE)$_2$ and of the He complexes. As for DFE monomer, the results are also of interest due to its aforementioned 'prototype' status and the fact that its spectrum has not previously been studied with high resolution in the mid-infrared region.

DFE is a planar molecule with C$_{2v}$ point group symmetry. Its structure is analogous to that of ethylene, but with two of the H atoms at one end replaced by F atoms. The other form, 1,2-DFE, has the two F atoms attached to different C atoms. Ground vibrational state (and some excited state) rotational parameters for DFE have been precisely determined using microwave and millimeter wave spectroscopy, beginning in 1957 with the work of Edgell et al. [3] and continuing with contributions from Laurie and Pence [4] and others. These results were summarized and extended in 1995 by Zerbe-Foese et al. [5], where further references can be found. Extensive work over the past 75 years on the vibrational structure of DFE is summarized and extended by Krasnoshchekov et al. [1] and McKean et al. [6].

The present spectra were recorded as described previously [7,8] using a pulsed supersonic jet and a tunable optical parametric oscillator or quantum cascade laser source. The expansion mixture normally contained about 0.1-0.3 % DFE in helium carrier gas, and the jet backing pressure was 13 atmospheres. The jet is produced by two side-by-side pulsed solenoid valves with multi-channel blocks attached to their bottom faces and a pair of 5 cm long adjustable slit jaws mounted on each block. The slits are aligned parallel to the axis of a 70 cm long multi-pass optical cell in which the laser beam makes approximately 100 passes through the region directly below the slit jet jaws. Data acquisition is based on a rapid-scan signal averaging mode with continuous background subtraction. The laser is rapidly and repeatedly scanned over a range of about 30 GHz (1 cm$^{-1}$). In the case of OPO the jet signal (idler) is being scaled and subtracted from output of an InGaAs detector monitoring the OPO signal channel and when using a QCL the jet signal is subtracted from a portion of the IR signal by passing the jet. These digitized sweep signals are averaged for hundreds of scans. Wavenumber calibration is carried out by simultaneously recording signals from a fixed etalon and a room temperature reference gas cell. Spectral simulations and fitting are made using the PGOPHER software [9].

## II.     The DFE dimer

We observed a number of bands which were plausible candidates to be due to DFE dimer, for which there are no previous spectroscopic results. Examples of these bands are shown in Fig. 1. Similar bands (not shown here) which were weaker, less well resolved, and/or more heavily overlapped by DFE monomer transitions were observed at 2253.29, 2643.34, 3053.06, and 3094.99 cm$^{-1}$.

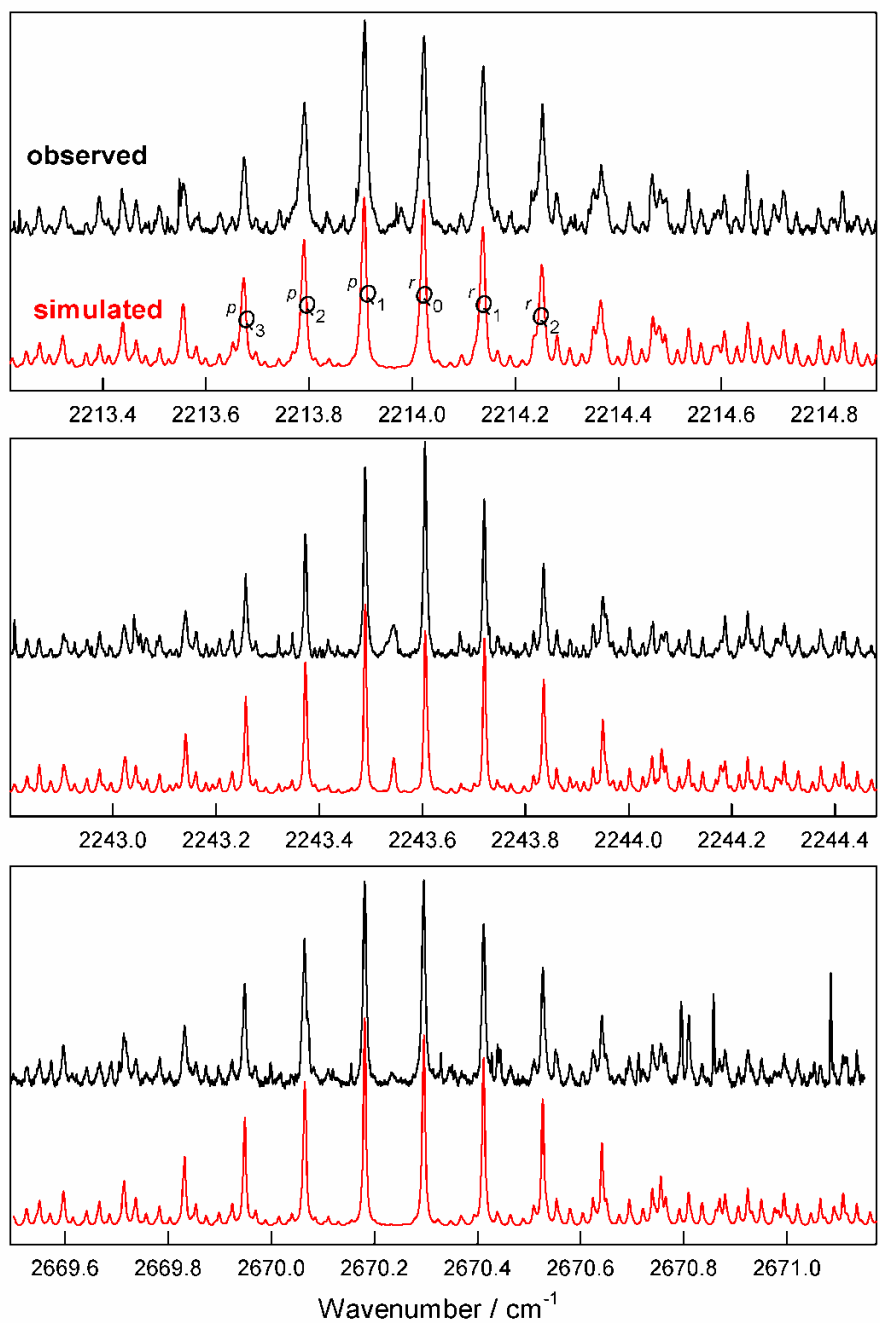

Fig. 1. Observed (in black) and simulated (in red) spectra of DFE dimer bands. The vertical scales of these spectra are multiplied by a factor of from 10 to 30 compared to the corresponding DFE monomer spectra in Section V.

In order to investigate the possible structure of DFE dimer, we performed *ab initio* structural calculations following the multi-step computational strategy we developed for studying gas-phase molecular clusters (a detailed description can be found in refs. [10,11]). Briefly, the minima on the potential energy surface of the DFE dimer were characterized by a composite procedure based on calculations carried out using different density functional levels of theory (DFT) of increasing complexity. Following our previous works [12], the final optimizations and subsequent frequency calculations were performed using two double-hybrid functionals, namely the rev-DSDPBEP86 [13] and B2PLYP [14]; besides, in line with the recent studies on fluoroethylene dimers [15], the geometry of the most stable isomer was carried out at the MP2 [16] levels of theory. In all the calculations the jun-cc-pVTZ basis set [17] was employed, and the Grimme's DFT-D3 corrections with Becke-Johnson damping [18,19] were included in all the DFT computations. Finally, equilibrium geometries were obtained at the rev-DSDPBEP86 level of theory, the jun-ChS extrapolation procedure [20] was carried out to provide accurate estimates of the corresponding binding energies. The Gaussian16 program [21] was employed for all the calculations, and the *Superfine* grid option was specified for the computations performed at DFT level of theory.

We identified four lowest energy isomers of DFE dimer (the corresponding equilibrium geometries and binding energies are reported in Table S1 of Supporting Information). The most stable calculated dimer structure locates the DFE monomers in a slipped antiparallel configuration with two-fold rotational symmetry around the $c$-inertial axis. As in the case of fluoroethylene dimers [15], the potential energy surface of DFE dimers appears to be so flat and slippery that even a very small energy difference may lead to significant effects in the optimized geometry. Actually, the geometry optimized at the B2PLYP level of theory has the four C atoms in the same plane, and the two monomer planes are parallel (thus in line with $C_{2h}$ symmetry), and also a similar structure is obtained at the B3LYP level of theory. At variance, both MP2 and rev-DSDPBEP86 calculations led to a different optimized geometry in which the two DFE monomers are slightly twisted with respect to each other ($C_2$ symmetry), as shown in Fig. 2, while the optimized structure having $C_{2h}$ symmetry is now characterized by a small negative frequency. But the energy barrier to a more symmetric $C_{2h}$ geometry (which has a plane of symmetry corresponding to the *a-b* plane) can be considered almost negligible (at the rev-DSDPBEP86 level of theory the calculated energy difference between the $C_2$ and $C_{2h}$ equilibrium structures is lower than 0.01 kcal mol$^{-1}$), so the effective dimer structure may well be $C_{2h}$.

The measured dipole moment of DFE monomer is 1.39 D [22] and our calculated value is 1.31 D (rev-DSDPBEP86). Our calculated dipole moment for the dimer is 0.31 D, and of course the effective dipole must be zero if the effective structure of the dimer is indeed $C_{2h}$. In any case, the results help to explain why no pure rotational spectrum has yet been observed for DFE dimer.

The calculated rotational parameters already explained the observed spectra rather well. These parameters were refined in a combined fit (see Table 1) of the five best resolved dimer bands, including a total of 255 observed lines with an overall root mean square (rms) deviation of 0.0010 cm$^{-1}$. Most of the lines were blends involving more than one unresolved transition, and we used the mergeblend option of PGOPHER to fit observed line positions to the intensity weighted average of the calculated transitions.

DFE dimer turns out to be very close to being an accidental symmetric top (the rev-DSDPBEP86 and B2PLYP calculations predicted a value of ($B - C$) equal to only 1.9 MHz and 2.0 MHz, respectively), so the spectra resemble perpendicular bands of a prolate symmetric rotor. Thus the observed spectra are dominated by unresolved perpendicular $Q$-branches, some of which are labeled in the top panel of Fig. 1. In our fits, the values obtained for the parameter ($B - C$) were essentially zero within experimental error (see Table 1). Thus we are not sure whether the axes indicated in Fig. 2 are correct, or whether *b* and *c* should be interchanged (there was a slight preference for the *c* as the symmetry axis, as in Fig. 2). Assuming two-fold rotational symmetry, as in the calculated structure, then the dimer will have nuclear spin weights of 136:120 for levels with $K_c$ = even:odd (or 136:120:120:136 for $K_a K_c$ = ee:eo:oe:oo if $b$ is the symmetry axis). However, it was not possible to verify this experimentally because the simulated spectra with and without spin weights were virtually identical.

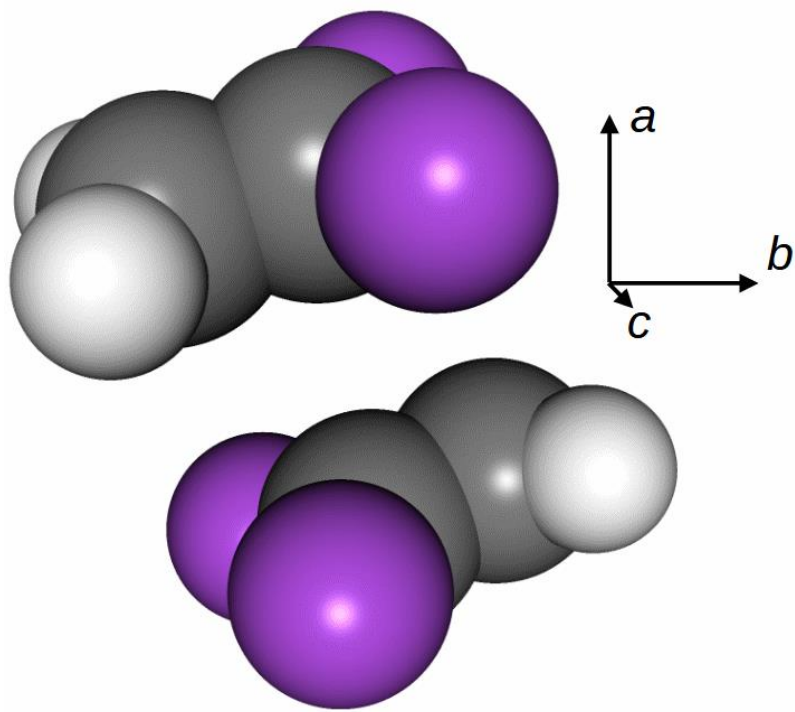

Fig. 2. Calculated equilibrium structure of 1,1-DFE dimer at rev-DSDPBEP86 level of theory. There is two-fold rotational symmetry around the $c$-inertial axis. In this view, the $c$-axis is almost, but not quite, perpendicular to the plane of the drawing. The DFE monomers are slightly twisted such that the four C atoms are not quite co-planar and the monomer planes are not quite parallel. But the tunneling barrier at the more symmetric transition state may be low enough to give the dimer effective C$_{2h}$ symmetry.

Table 1. Molecular parameters for the observed bands of DFE dimer (in cm$^{-1}$). [a]

| $\nu_0$ | $A$ | $(B+C)/2$ | $(B-C)$ | width | Monomer vibration | Sym | $\Delta\nu$ | $n$ | rms |
|---|---|---|---|---|---|---|---|---|---|
| ground (calc)[b] | 0.092582 | 0.036950 | 0.000032 | | | | | | |
| ground (fit) [c] | 0.093163(16) | 0.0352932(95) | 0.000035(20) | | | | | | |
| 2213.9672(2) | 0.092940(16) | 0.035217(11) | 0.000020(56) | 0.007 | 2219.660 | B$_1$ | -5.692 | 72 | 0.0013 |
| 2243.5461(2) | 0.092994(16) | 0.035320(10) | 0.000091(41) | 0.0035 | 2250.273 | A$_1$ | -6.727 | 82 | 0.0007 |
| 2253.2934(4) | 0.093161(47) | 0.035137(16) | 0.000002(54) | 0.005 | ? | | | 24 | 0.0008 |
| 2643.3435(5) | 0.092802(47) | 0.035290(16) | 0.00009(11) | 0.012 | 2648.113 | A$_1$ | -4.769 | 27 | 0.0011 |
| 2670.2391(3) | 0.093011(22) | 0.035281(13) | 0.000068(52) | 0.005 | 2673.863 | B$_1$ | -3.624 | 50 | 0.0007 |
| 3053.06[d] | | | | 0.03 | 3058.102 | A$_1$ | -5.04 | | |
| 3094.99[d] | | | | 0.03 | 3098.498[e] | A$_1$ | -3.51 | | |

[a] width is the approximate additional width presumed due to predissociative broadening. Monomer vibration is the DFE monomer band (see Section V) most likely associated with the dimer band, and Sym is the symmetry of that monomer band. $\Delta\nu$ is the vibrational frequency shift relative to the associated DFE monomer band. $n$ is the number of measured line positions in the fit for that band. Rms is the root mean square deviation of the fit for that band.

[b] Computed from the equilibrium geometry at the rev-DSDPBEP86 level of theory (see text).

[c] The fitted ground state parameters determined in the combined fit of five dimer bands.

[d] The 3053.06 and 3094.99 cm$^{-1}$ bands were too weak and broadened to fit in detail.

[e] Or 3101.368 cm$^{-1}$, in which case $\Delta\nu$ = -6.38 cm$^{-1}$.

For the purpose of discussion, we assume here that *c* is the symmetry axis and that the dimer has effective $C_{2h}$ symmetry. Results would be similar if there were a small deviation from $C_{2h}$. Each monomer vibrational band gives rise to two dimer bands, corresponding to in-phase or out-of-phase vibrations on the two equivalent monomers. Because DFE monomer has $C_{2v}$ point group symmetry, see Section V, its excited states with $A_1$ symmetry give rise to bands with *a*-type rotational selection rules, those with $B_1$ symmetry give rise to *b*-type transitions, and states with $B_2$ symmetry give rise to *c*-type transitions. This means that if the monomer vibrational transition is *a*-type, then the in-phase dipole transition moments in the dimer cancel and there is no transition, while the out-of-phase moments add constructively to give a mostly *b*-type dimer transition with a small *a*-type component. The relative dipole projection on the *a*-axis is 29% in our calculated structure, so the *a*-type component would have about 8% of the band strength. If the monomer transition is *b*-type, then the out-of-phase dipoles in the dimer cancel while the in-phase dipoles give an entirely *c*-type transition.

In the observed spectra, it was clear that the 2243.546 cm$^{-1}$ dimer band had an *a*-type component (see the center panel of Fig. 1), which suggests that it is associated with the $A_1$ 2250.273 cm$^{-1}$ monomer vibration, and not the $B_1$ 2240.061 cm$^{-1}$ vibration as we first thought (see Section V for the monomer results). But this leaves a problem of what monomer vibration to associate with the 2253.293 cm$^{-1}$ dimer band, and we do not have a solution for this. There is also an apparent *a*-type component for the 2643.344 cm$^{-1}$ dimer band, so it can appropriately be associated with the $A_1$ 2648.113 cm$^{-1}$ monomer vibration. There is no apparent *a*-type component for the 2213.967 cm$^{-1}$ dimer band, so it can appropriately be associated with the $B_1$ 2219.660 cm$^{-1}$ monomer vibration (a small peak near the band center is not in the right position to be the *a*-type *Q*-branch). The presence or absence of *a*-type components for the remaining dimer bands is not very clear, and the most likely associated monomer bands are shown in Table 1. It is interesting to note that all the dimer bands are red-shifted from their monomer partners by roughly similar amounts ranging from 3.5 to 6.7 cm$^{-1}$.

If the DFE dimer symmetry is $C_2$, then five parameters are required to specify the dimer structure (assuming unchanged monomers) and a purely experimental structure determination is not possible. If however there is effective $C_{2h}$ symmetry (as we suspect considering also the results obtained at the B3LYP and B2PLYP levels of theory) then only two parameters are required. Using the three observed ground state rotational constants, a fairly good fit is obtained with structural parameters $R = 3.44$ Å for the separation of the monomer centers of mass, and $\theta = 83.9°$ for the angle C1-C2-C3 (if C1 is the F-end carbon atom of monomer 1, then C2 is the other C atom of monomer 1 and C3 is the F-end carbon atom of monomer 2; $\theta = 90°$ would mean that the four C atoms form a rectangle). Another way to specify this structure is with the two inter-monomer C atom separations, which are 3.43 and 3.38 Å (F-end to H-end, and F-end to F-end, respectively). For comparison, the geometry optimized at the rev-DSDPBEP86 level of theory has these separations equal to 3.43 and 3.30 Å (they become 3.44 and 3.35 Å in the case of the structure obtained using the B2PLYP functional).

## III.    DFE-He

Weak unexplained lines were present in some of the stronger DFE monomer bands, and they were particularly noticeable close to the band origins. Figure 3 shows a magnified view of the central region of the 2219.660 cm$^{-1}$ monomer band. See Section V. The top trace in Fig. 3 (in black) uses a more dilute gas mixture ($\approx 0.1$ % DFE in helium) and the weak lines became relatively stronger, supporting the idea that they were due to DFE-He. In addition, some new weak lines appeared which we assigned to DFE-He$_2$. A more extended view of the DFE-He and -He$_2$ transitions is shown in Fig. 4, in which the much stronger DFE monomer lines have been cut out for clarity.

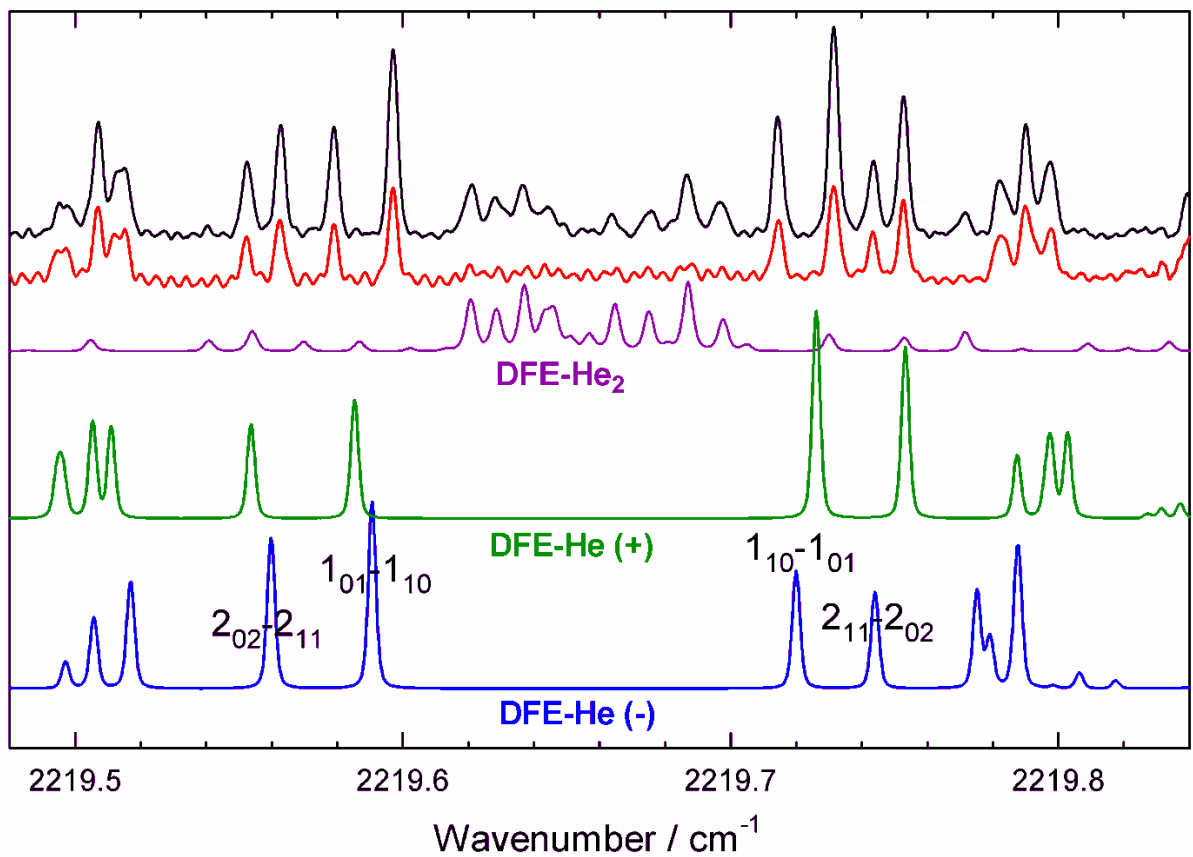

Fig. 3. Expanded view near the center of the 2219.66 cm⁻¹ DFE monomer band, showing observed (upper two) and simulated (lower three) spectra of DFE-He and DFE-He₂. The observed spectrum in red uses a higher concentration of DFE while the spectrum in black was obtained with a more dilute mixture of DFE in helium, enhancing DFE-He and greatly enhancing DFE-He₂. Rotational assignments are given for some DFE-He transitions.

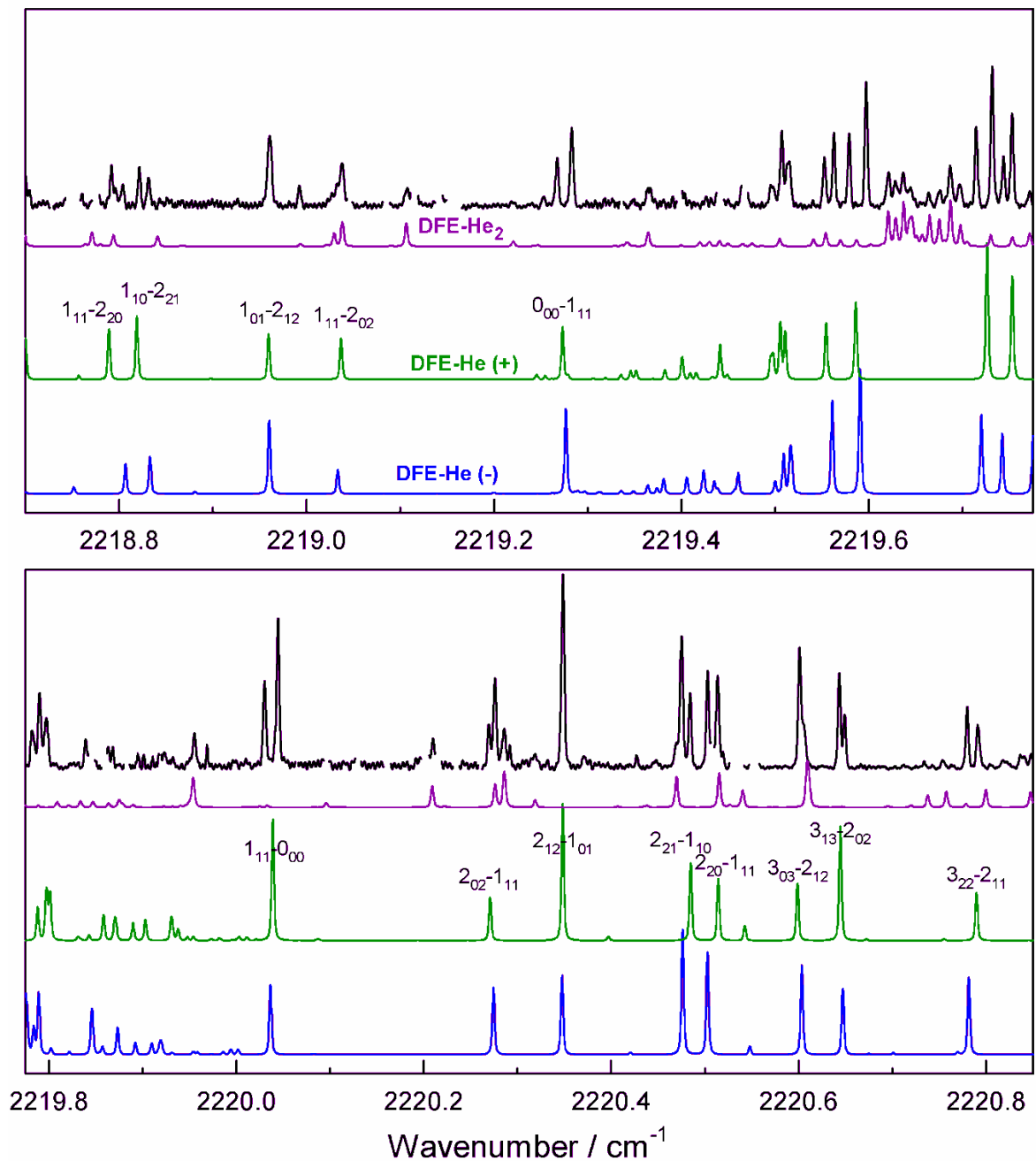

Fig. 4. Extended view of the 2219.66 cm⁻¹ DFE-He and DFE-He₂ bands. The observed spectrum (black) is the same as the upper trace in Fig. 3. For clarity, the strong DFE monomer lines have been cut out, resulting in small blank spaces in the observed spectrum. Rotational assignments are given for stronger DFE-He transitions; see Fig. 3 for more assignments in the central region.

Based on previous studies of DFE-Ne and -Ar [23,24], the structure of DFE-He is very likely to be similar to that shown here in Fig. 5, with the He atom located out-of-plane just beyond the F end of DFE. Using the structure determined for DFE-Ne but substituting He, we obtained estimated rotational parameters which approximately explained the spectrum, but the predicted lines were usually observed to be doubled. This was not surprising since small doubling effects (<1 MHz) were previously observed in the millimeter wave spectrum of DFE-Ne [24] and it was natural to expect much larger effects for DFE-He. This doubling corresponds to tunneling of He (or Ne) from one side of the DFE to the other through a planar $C_{2v}$ transition state. To visualize the doubling in the spectrum, take for example the four lines observed in Fig. 3 from 2219.54 to 2219.60 cm$^{-1}$. In the absence of tunneling we expect two transitions here ($1_{01}\leftarrow 1_{10}$ and $2_{02}\leftarrow 2_{11}$), but four lines are observed. The lowest symmetric (+) tunneling level has $A_1$ symmetry in $C_{2v}$, and its nuclear spin weights are 10:6 for levels with $K_a$ = even:odd, same as for DFE monomer. The upper antisymmetric (–) tunneling level has $B_2$ symmetry, and its weights are reversed to 6:10 for $K_a$ = even:odd. The effect of the weights is readily observable in Fig. 3, where the (–) tunneling lines are stronger than the corresponding (+) lines at 2219.54 to 2219.60 cm$^{-1}$, while the opposite is true for the mirror image transitions ($1_{10}\leftarrow 1_{01}$ and $2_{11}\leftarrow 2_{02}$) at 2219.70 to 2219.76 cm$^{-1}$. The magnitude of the splitting is not constant because the (+) and (–) states have different rotational constants. In some cases, doubling was not resolved because the two lines coincided within the experimental line width (for example, the line at 2218.96 cm$^{-1}$ in Fig. 4). When splitting was resolved, as it usually was for reasonably isolated lines, the assignments to (+) and (–) were easily chosen thanks to the different spin weights.

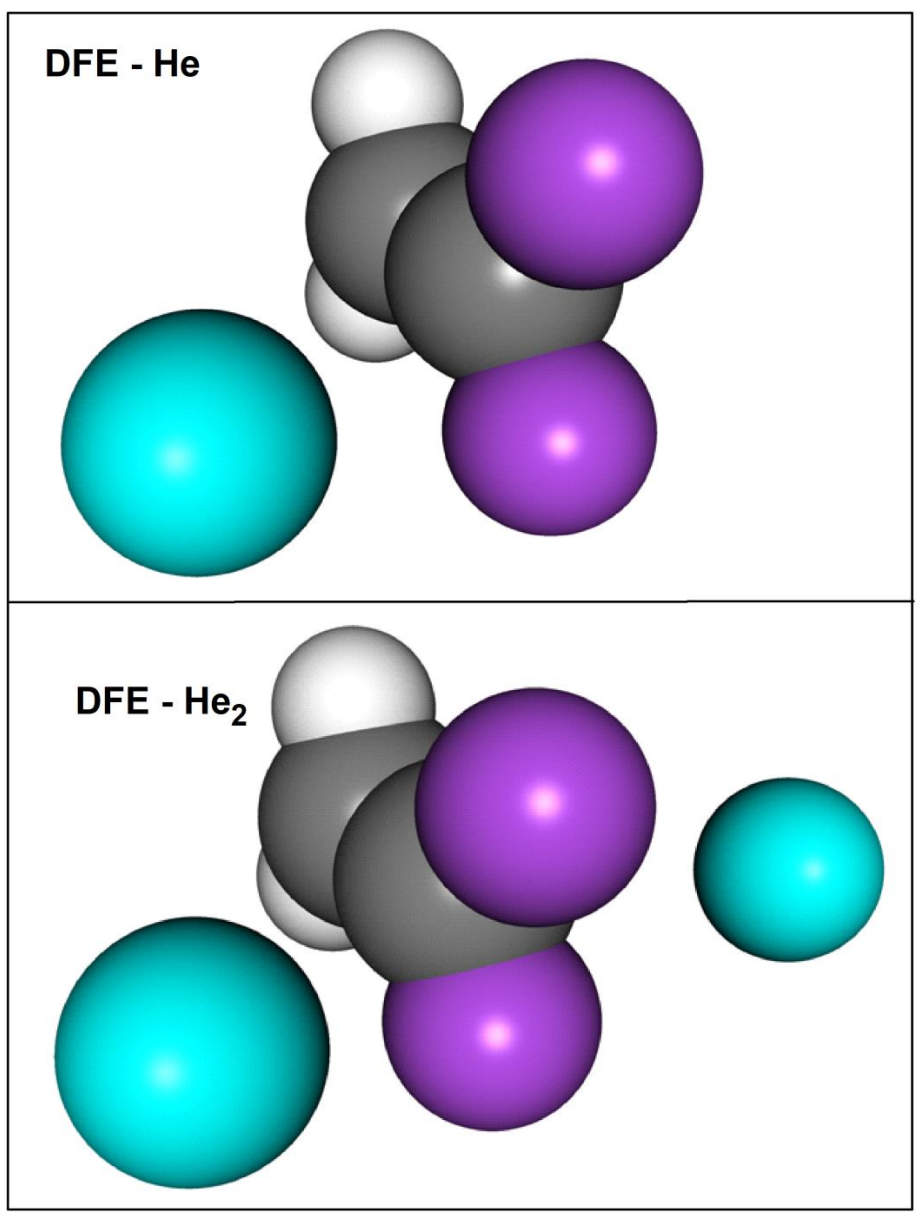

Fig. 5. Structures of DFE-He and DFE-He$_2$. For DFE-He, tunneling of the He atom through the DFE monomer plane (C$_{2v}$ transition state) gives rise to splittings in the spectrum. For DFE-He$_2$, the second He atom occupies a position equivalent to the first, and tunneling is quenched. For DFE-He, the experimental structure gives 3.424 Å for distance between the DFE and He centers of mass, and 68.9° for the angle between the DFE monomer symmetry axis and the line connecting these centers of mass. For DFE-He$_2$, the corresponding values are 3.458 Å and 72.5°.

Table 2. Molecular parameters for the observed bands of DFE-He (in cm$^{-1}$). [a]

| | $\nu_0$ | $\Delta\nu$ | $A$ | $B$ | $C$ | $D_{JK}$ | $n$ | rms |
|---|---|---|---|---|---|---|---|---|
| (+) | ground | | 0.22736(65) | 0.17918(45) | 0.15634(31) | 0.000134(19) | | |
| (−) | ground | | 0.22165(64) | 0.17845(54) | 0.15699(30) | 0.000034(18) | | |
| (+) | 2219.6572(15) | -0.0023 | 0.22557(55) | 0.17909(51) | 0.15532(32) | 0.000163(17) | 29 | 0.0026 |
| (−) | 2219.6567(16) | -0.0028 | 0.22137(52) | 0.17759(52) | 0.15624(33) | 0.000105(15) | 27 | 0.0028 |
| (+) | 2250.2521(12) | -0.0205 | 0.22616(50) | 0.17899(26) | 0.15637(22) | [0.0001] | 14 | 0.0021 |
| (−) | 2250.2526(12) | -0.0200 | 0.22136(56) | 0.17831(23) | 0.15690(17) | [0.0001] | 15 | 0.0020 |
| (+) | 2592.9362(19) | +0.0051 | 0.22611(37) | 0.17840(39) | 0.15565(41) | [0.0001] | 14 | 0.0021 |
| (−) | 2592.9369(19) | +0.0055 | [0.221] | 0.17763(40) | 0.15654(37) | [0.0001] | 9 | 0.0027 |
| (+) | 2648.1488(11) | +0.0357 | 0.22575(41) | 0.17805(18) | 0.15535(25) | [0.0001] | 19 | 0.0021 |
| (−) | 2648.1503(11) | +0.0367 | 0.21962(13) | 0.17748(24) | 0.15625(29) | [0.0001] | 15 | 0.0023 |
| (+) | 2673.9132(23) | +0.0502 | 0.22613(54) | 0.17825(96) | 0.15525(34) | [0.0001] | 12 | 0.0032 |
| (−) | 2673.9088(18) | +0.0458 | 0.22247(24) | 0.17800(82) | 0.15669(34) | [0.0001] | 14 | 0.0033 |
| (+,−) | 3052.309(4) | +0.031 | | | | | | |
| (+,−) | 3058.098(3) | -0.004 | | | | | | |

[a] (+) and (−) represent the symmetric and antisymmetric tunnelling components. $\Delta\nu$ is the vibrational frequency shift relative to the DFE monomer. The ground state parameters were determined in the fit to the 2219.66 cm$^{-1}$ band, and then held fixed in the other band fits. Square brackets indicate that the parameter was held fixed at the indicated value.

DFE-He spectra were detected accompanying seven of the strongest DFE monomer bands. Selection rules were $a$-type for monomer bands with $A_1$ upper states and $b$-type for those with $B_1$ upper states. Our rotational analysis focused on the 2219.66 cm$^{-1}$ band for which the best DFE-He (and DFE-He$_2$) results were obtained, and the top of Table 2 shows the ground and excited state rotational parameters obtained from a fit to this band. These ground state parameters were then held fixed in less complete rotational fits to four other bands (2250.27, 2592.93, 2648.11, 2673.86 cm$^{-1}$), while for two bands (3052.26, 3058.10 cm$^{-1}$) we could only estimate band origins. The $\Delta\nu$ values in Table 2 represent vibrational frequency shifts of the DFE-He band origins relative to the free DFE monomer. Both positive (blue-shift) and negative (red-shift) values were observed, and all are quite small in magnitude (<0.04 cm$^{-1}$). The DFE-He line widths were difficult to determine, especially for the bands associated with $A_1$ monomer vibrations, where the DFE-He lines tend to overlap each other. However, they all tended to be slightly broader than those of the DFE monomer, with approximate excess broadenings of 0.001 cm$^{-1}$ for the 2219.66 and 2673.91 cm$^{-1}$ bands, and 0.002 cm$^{-1}$ for the 2250.25 and 2592.94 cm$^{-1}$ bands. The broadening is presumably due to predissociation, that is, the finite lifetimes of the upper states. There did not seem to be any obvious correlation with the DFE dimer widths (Table 1).

The rms deviation in fitting 56 observed lines of the 2219.66 cm$^{-1}$ band in terms of 18 parameters was 0.0027 cm$^{-1}$, which is not particularly good! This problem is evident in Fig. 3, where the simulated DFE-He transitions do not match the observed spectrum particularly well. This unreliability meant that assignments were often not possible in certain regions of the spectrum with many partially overlapping lines. Similarly imperfect fits were also observed for the other DFE-He bands. The problem was not simply due to perturbations in the excited states (though these may also exist), and the poor fits could not simply be explained by centrifugal distortion or in terms of the possible $b$-type Coriolis interaction between the (+) and (–) states. A systematic trend was noted in the sense that the (+) and (–) state fitting residuals (obs – calc) tended to be opposite and roughly equal for a given transition, suggesting that a better fit could be obtained using averages of the (+) and (–) line pairs. When we tried an averaged fit, the resulting parameters were as expected about midway between those of the separate (+) and (–) fits, and the rms deviation was reduced by a factor of almost 3. The tunneling splittings are of course not constant and the implication seems to be that they are not well modeled by the rotational energy expression. Could this effect be related to the possibility of other He atom tunneling paths (in addition to simply going between the F atoms)? In any case, we think that the conventional semi-rigid molecule Hamiltonian is not fully adequate in this case due to these large amplitude multi-dimensional He atom motions. This is not unexpected since conventional rotational energy expressions are also inadequate for other He-containing van der Waals complexes including CO-He [25], $N_2O$-He [26], OCS-He [27], and many more.

The ground state $A$ rotational constant of the DFE-He (+) component is significantly larger (2.5%) than that of the (–) component. This is the expected trend since the (–) tunneling wavefunction passes through zero at the $C_{2v}$ tunneling barrier where $A$ has a maximum, whereas the (+) wavefunction is finite. In other words, on average, the He atom tends to be located closer to the monomer symmetry axis for the (+) component. The differences between the (+) and (–) band origins in Table 2 are almost negligible (<0.0008 cm$^{-1}$) in three cases and very small (<0.005 cm$^{-1}$) in two others, meaning that there is very little change in the rotationless tunneling splitting between the ground and excited DFE intramolecular vibrational states. We do not have a direct measure of this splitting for the ground state, but the observed line intensities may give some information. Taking 8 strong isolated line pairs and correcting for the spin weights, we find on average that the (+) lines are (1.07 ± 0.06) times stronger than the (–) lines. Assuming a Boltzmann distribution, this suggests that the tunneling splitting is very likely smaller than 0.2 cm$^{-1}$ (perhaps about 0.1 cm$^{-1}$).

To obtain an idea of the approximate DFE-He structure, we fit the observed rotational constants and obtained effective values of ($R$, $\alpha$) = (3.424 Å, 68.9°) for the (+) tunneling component and (3.453 Å, 70.6°) for the (–) component. Here $R$ is the distance from the He atom to the DFE monomer center of mass, and $\alpha$ is the angle between the $R$ vector and the monomer symmetry axis. For comparison, these structural parameters are (3.29 Å, 67.8°) for DFE-Ne, and (3.51 Å, 71.2°) for DFE-Ar [23]. The fact that $R$ for DFE-He is smaller than for DFE-Ar but larger than for DFE-Ne illustrates a common trend in weakly-bound complexes. Lighter rare gas atoms are smaller, so $R$ values drop in going from Xe to Ne, but He is so light that large amplitude anharmonic effects ($R_0 > R_e$) tend to dominate. Thus, for example, the $R$-values of Rg-CO complexes vary as (4.08, 3.65, 3.85, 3.98, 4.19 Å) for the series (He, Ne, Ar, Kr, Xe) [28].

## IV.    DFE-He$_2$

For experimental reasons, we only obtained a clear spectrum of DFE-He$_2$ in the region of the 2219.66 cm$^{-1}$ monomer band (see Figs. 3 and 4). It is natural to assume that the second He atom in DFE-He$_2$ is located in a position equivalent to the first on the other side of the DFE monomer, as shown in Fig. 5. This geometry explains the observed spectrum well. DFE-He$_2$ has C$_{2v}$ symmetry, but its inertial axes are flipped compared to DFE monomer: $b$ is now the two-fold rotational symmetry axis (i.e. the C-C axis), $a$ is now perpendicular to the plane of the monomer (i.e. parallel to the He-He axis), and $c$ now lies in the plane of the monomer (parallel to the F-F axis). Thus the DFE-He$_2$ nuclear spin statistical weights for the vibrational ground state are 10 for ($K_a$, $K_c$) = (even, even) or (odd, odd) and 6 for (even, odd) or (odd, even). This intensity alternation is indeed evident in the spectrum, though not as clearly as for DFE-He. Whereas the 2219.66 cm$^{-1}$ band of DFE-He is $b$-type, that of DFE-He$_2$ is $a$-type due to the axis switching. This has the fortunate effect of separating the two spectra (at least partly) in the central region (see Fig. 3).

Table 3. Molecular parameters for the 2219.66 cm$^{-1}$ band of DFE-He$_2$ (in cm$^{-1}$). [a]

| | $\nu_0$ | $\Delta\nu$ | $A$ | $B$ | $C$ |
|---|---|---|---|---|---|
| ground | | | 0.16699(27) | 0.12797(17) | 0.11658(30) |
| | 2219.6603(8) | +0.0007 | 0.16620(18) | 0.12752(15) | 0.11633(30) |

[a] $\Delta\nu$ represents the vibrational frequency shift relative to the DFE monomer.

The transitions of DFE-He$_2$ were relatively weak and often overlapped with each other or obscured by those of DFE monomer and DFE-He, so our analysis was limited to only 27 lines. The quality of the fit was not particularly good (rmsd = 0.0019 cm$^{-1}$), but the parameters (shown in Table 3) seem reasonable (for example, the rotational constants decrease by 2 to 5% in the excited state, similar to those of DFE-He for this band). Unlike DFE-He, there was no obvious sign of systematic failure of the rotational Hamiltonian. But stronger and better resolved spectra could possibly show tunneling effects, which in this case might involve internal rotation of the He$_2$ unit relative to the DFE unit around the monomer symmetry axis. Fitting the ground state rotational constants gave a structure with ($R$, $\alpha$) = (3.458 Å, 72.5°), which is not so different from DFE-He. The slight increase in each structural parameter compared to DFE-He means that the He atoms in DFE-He$_2$ are located further from the symmetry axis than in DFE-He. This is presumably a consequence of the absence of tunneling through the DFE plane and the presence of He-He repulsion in DFE-He$_2$. One might expect the vibrational shift of DFE-He$_2$ to be double that of DFE-He, namely about -0.005 cm$^{-1}$, but this is not the case. Instead, the shift for DFE-He$_2$ is effectively zero (within experimental uncertainty), which must be related to the slightly different effective location of the He atoms relative to the DFE monomer.

**V.     The DFE monomer**

DFE monomer is a planar near-oblate asymmetric rotor, with rotational parameters $A \approx B \approx 2C$. The $C_2$ rotational axis coincides with the *a*-inertial axis, and the resulting ground state nuclear spin weights are 10:6 for levels with $K_a$ = even:odd. Three possible types of rotation-vibration bands can originate from the ground vibrational state. Excited states with $A_1$ symmetry give rise to *a*-type bands, states with $B_1$ symmetry give rise to *b*-type transitions, and states with $B_2$ symmetry give rise to *c*-type transitions. Meanwhile, transitions from the ground state to $A_2$ excited states are forbidden. Convention often chooses *c* to be the $C_{2v}$ x-axis [9]. However, here we follow most previous work on DFE [1,5,6], which makes the alternate choice, with *b* as the $C_{2v}$ x-axis so that $B_1$ states have *b*-type selection rules.

We ultimately observed a total of 23 bands of 1,1-DFE. These are summarized in Table S2. Our coverage of the spectral region from 2210 to 3105 cm$^{-1}$ was not continuous. Rather we jumped between different regions looking for strong and interesting bands, partly in an effort to find clear bands of (DFE)$_2$ and DFE-He. About half of the observed DFE monomer bands were reasonably strong, with well-resolved rotational transitions which could be easily assigned and analyzed. The remaining bands were then mostly detected in the process of trying to identify weak unexplained lines among the stronger bands. One vibration (2651.502 cm$^{-1}$) was only detected by means of its perturbation of a stronger band, and three very weak bands turned out not to originate from the ground state, but rather were hot bands originating in the lowest excited vibration, $\nu_{10}$, which is located at 436.885 cm$^{-1}$ [5]. All of the observed bands were either *a*- or *b*-type, and these both have similar appearances since $A \approx B$. The bands are dominated by groups of closely-spaced transitions separated by $2C \approx A \approx B \approx 0.35$ cm$^{-1}$, and the spacings within each group depend on the value of $(A - B)$ and on whether it is *a*- or *b*-type. Each group is not simply a $P(J)$, $Q(J)$, or $R(J)$ multiplet, but rather each has common values of $K_c$ for $Q(J)$ transitions, or of $(2J - K_c)$ for $P(J)$ and $R(J)$ transitions. The rotational assignments and fitting were straightforward, and the resulting molecular parameters are given in Table S2. Some examples of the spectra are shown in Figs. 6 and 7. Table S2 includes estimates of the relative intensities of the various DFE monomer bands, whose reliability is suggested by the fact that they are given with only two significant figures. Of course the relative intensities for nearby bands will be more accurate than those for more distant bands, for which various experimental conditions may have changed.

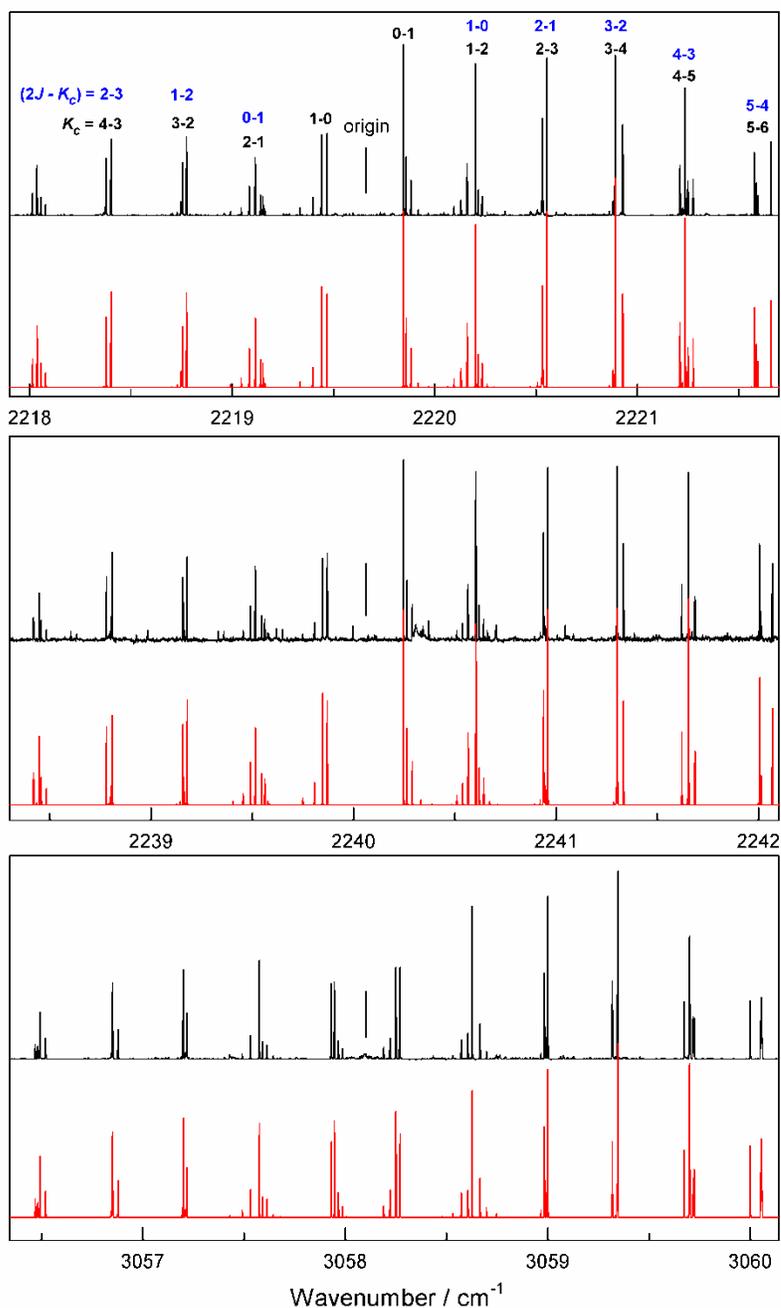

Fig. 6. Observed (in black) and simulated (in red) spectra of some 1,1-DFE monomer bands. Band origins are aligned at the positions indicated by the vertical lines. The 2219.660 and 3058.102 cm⁻¹ bands are *a*-type, and the 2240.061 cm⁻¹ band is *b*-type. See Tables 4 and 5 for vibrational assignments. The simulations assume an effective rotational temperature of 2.3 K and a Gaussian line width of 0.0022 cm⁻¹. Closely-spaced groups of lines are spaced by $2C \approx 0.35$ cm⁻¹. Each member of a group has common values of $K_c$ for $Q$ transitions, or of $(2J - K_c)$ for $P$ and $R$ transitions, as indicated in the top panel.

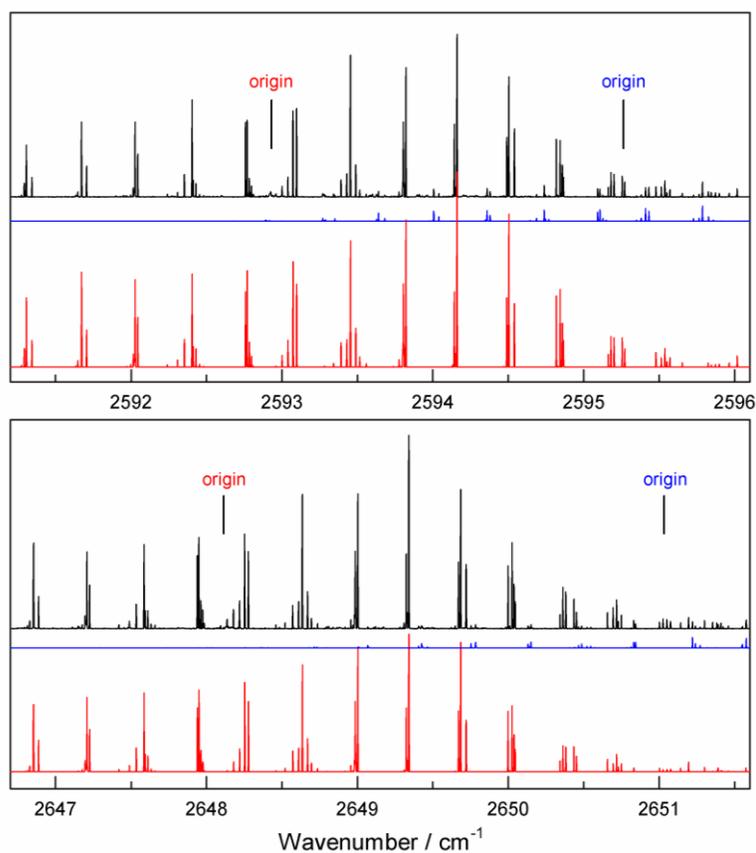

Fig. 7. Further examples of DFE monomer spectra (observed in black, simulated in color). In both cases shown here a strong band (simulated in red) is accompanied by a weaker band (in blue) at higher wavenumber. The 2648.113 cm$^{-1}$ band, assigned as $\nu_2 + \nu_4$, was the strongest of all those observed in the present study, and the 2592.931 cm$^{-1}$ band, assigned as $2\nu_8$, was the third strongest.

The low effective rotational temperature ($\approx$2.3 K) in the supersonic jet means that only fairly low rotational levels were observed (generally $J < 11$, $K_a < 6$). However, this temperature was not particularly well defined, in the sense that transitions arising from higher energy levels were progressively stronger than would be expected based on the temperature that fitted lower energy levels. This sort of non-Boltzmann distribution is a normal effect in supersonic jets, and is due to less efficient rotational relaxation for higher and more widely spaced energy levels. It was more pronounced here for DFE monomer than in our usual studies of weakly-bound clusters. The present spectra are of course simpler and much easier to analyze compared to DFE monomer spectra recorded at room temperature (or even chilled to 142 K as in [5]). The present results, though incomplete, demonstrate that valuable new vibrational data for a "medium size" molecule can be obtained by combining a tunable infrared laser source with a dilute helium supersonic jet.

In the present analyses, rotational constants for the ground state (and the $v_{10} = 1$ state) were fixed at the values reported by Zerbe-Foese et al. [5], including quartic and sextic distortion parameters. For the excited states, quartic parameters were fixed at the same lower state values and sextic parameters were fixed to zero (this latter choice made absolutely no difference to the quoted results because of the low $J$-values involved). The band origins quoted in Table S2 provide new information for vibrational studies such as that in [1], and the excited state rotational parameters should be useful starting points for future high-resolution studies of these bands at higher temperatures.

Vibrational assignments for our 23 observed bands were not always straightforward. The density of DFE monomer vibrational states increases steadily with increasing wavenumber, so that there is (very roughly) about one state for each 3 cm$^{-1}$ around 2000 cm$^{-1}$ and almost double this density around 3000 cm$^{-1}$. The division into four distinct vibrational symmetries ($A_1$, $A_2$, $B_1$, $B_2$) helps to reduce the resulting confusion, but the Fermi-type interactions among states within each symmetry type still cause the exact labeling of a given state to become increasingly ambiguous. In the following discussion, it should be kept in mind that there are no completely "pure" vibrational assignments.

Table 4. Vibrational assignments for stronger DFE monomer bands, based on Table 7 of [1] (in cm$^{-1}$). [a]

| | observed (present) | int | assignment [1] | observed [6] | calculated [1] | sum of fundamentals |
|---|---|---|---|---|---|---|
| $B_1$ | 2219.660 | 0.65 | $\nu_4 + \nu_8$ | 2220.2 | 2219.3 | 2227.2 |
| $A_1$ | 2250.273 | 0.60 | $\nu_8 + \nu_9$ | 2250.3 | 2250.3 | 2255.3 |
| $A_1$ | 2592.931 | 0.75 | $2\nu_8$ | 2592.6 | 2593.6 | 2602.8 |
| $A_1$ | 2648.113 | 1.2 | $\nu_2 + \nu_4$ | 2647.6 | 2647.2 | 2654.1 |
| $B_1$ | 2673.863 | 0.3 | $\nu_2 + \nu_9$ | 2674.0 | 2672.6 | 2682.2 |
| $A_1$ | 2777.699 | 0.08 | $2\nu_3$ [b] | 2777.7 | 2774.2 | 2717.8 |
| $B_1$ | 3016.618 | 0.09 | $\nu_2 + \nu_8$ | 3015.5 | 3017.1 | 3029.7 |
| $B1$ | 3050.961[c] | 0.017 | $\nu_5 + \nu_8 + 2\nu_{12}$ | | 3052.4[c] | 3070.2 |
| $A_1$ | 3058.102 | 1.0 | $\nu_1$ | 3057.8 | 3057.8 | 3057.8 |
| $A_1$ | 3098.498 | 0.23 | $\nu_3 + \nu_6 + \nu_{10} + \nu_{12}$ | 3098.6 | 3096.2 | 3113.6 |
| $A_1$ | 3101.368 | 0.40 | $\nu_2 + \nu_9 + \nu_{10}$ | 3101.4 | 3099.9 | 3119.2 |

[a] int is the approximate relative intensity. 'Sum of fundamentals' is a simple sum of the indicated fundamental wavenumbers, which are taken from Table 4 of [1].

[b] According to [1] this state also contains a large contribution from $2\nu_6 + \nu_{11} + \nu_{12}$.

[c] See the text for a discussion of this weak $B_1$ band and its strong $A_1$ companion at 3052.260 cm$^{-1}$ (Table 5).

Table 5. Possible vibrational assignments for the remaining DFE monomer bands (in cm$^{-1}$).

| | observed (present) | int | assignment | sum of fundamentals |
|---|---|---|---|---|
| A$_1$ | 2216.582 | 0.01 | $\nu_3 + 2\nu_{10}$ | 2233.1 |
| B$_1$ | 2240.061 | 0.15 | $\nu_4 + 3\nu_{10}$ | 2237.1 |
| | | | $\nu_4 + \nu_6 + \nu_{12}$ | 2243.5 |
| A$_1$ | 2246.497 | 0.01 | $\nu_9 + 3\nu_{10}$ | 2265.2 |
| | | | $\nu_2 + \nu_5$ | 2278.1 |
| A$_1$ | 2595.265 | 0.09 | $\nu_2 + 2\nu_{10}$ | 2602.5 |
| | | | $\nu_9 + \nu_{10} + 2\nu_{12}$ | 2610.0 |
| | | | $\nu_8 + 3\nu_{10}$ | 2612.7 |
| B$_1$ | 2651.033 | 0.11 | $\nu_3 + \nu_8$ | 2660.3 |
| A$_2$[a] | 2651.502 | 0.0 | $\nu_{10} + 2\nu_{11} + \nu_{12}$ | 2650.8 |
| | | | $\nu_5 + \nu_8 + \nu_{11}$ | 2653.3 |
| | | | $2\nu_6 + \nu_{10} + \nu_{11}$ | 2655.6 |
| | | | $2\nu_5 + \nu_9 + \nu_{12}$ | 2663.0 |
| B$_1$ | 2673.026 | 0.02 | $\nu_3 + 3\nu_{10}$ | 2670.2 |
| | | | $\nu_3 + \nu_6 + \nu_{12}$ | 2676.6 |
| | | | $\nu_8 + \nu_9 + \nu_{10}$ | 2692.4 |
| A$_1$ | 3047.840 | 0.012 | $3\nu_5 + \nu_9 + \nu_{10}$ | 3040.4 |
| A$_1$ | 3052.260 [b] | 0.30 | $3\nu_5 + \nu_{11} + \nu_{12}$ | 3061.0 |
| | | | $3\nu_5 + 2\nu_6$ | 3065.8 |
| | | | $2\nu_4 + 2\nu_{12}$ | 3070.6 |
| | | | $4\nu_5 + 2\nu_{10}$ | 3073.4 |
| B$_1$[c] | 2689.932 | | $\nu_8 + \nu_9 + \nu_{10}$ | 2692.4 |
| B$_1$[c] | 3080.820 | | $\nu_2 + \nu_4 + \nu_{10}$ | 3091.2 |
| B$_1$[c] | 3533.479 | | $\nu_3 + \nu_6 + 2\nu_{10} + \nu_{12}$ | 3550.8 |

[a] This state is observed only via its perturbation of 2651.033 band.

[b] See the text for a discussion of this A$_1$ band and its weak B$_1$ companion at 3050.961 cm$^{-1}$ (Table 4).

[c] These are hot bands originating in the $\nu_{10} = 1$ state, which is located at 436.8851 cm$^{-1}$. The band origins are 2253.0467(2), 2643.9648(1) and 3096.6238(2) cm$^{-1}$, and the presumed hot band assignments are $\nu_8 + \nu_9 + \nu_{10} - \nu_{10}$, $\nu_2 + \nu_4 + \nu_{10} - \nu_{10}$ and $\nu_3 + \nu_6 + 2\nu_{10} + \nu_{12} - \nu_{10}$.

DFE monomer vibrational states in the 400 to 3500 cm$^{-1}$ range have been analyzed and assigned in detail by Krasnoshchekov et al. [1] using a numerical-analytic operator version of canonical van Vleck perturbation theory. Their Table 7 represents the best current state of knowledge of this subject and also demonstrates the extensive mixing mentioned above. Among our 22 observed bands, there are 11 which can be assigned with reasonable confidence on the basis of their table. These are listed here in Table 4, which repeats the observed band origins, symmetries, and intensities from Table S2. It also includes previously observed [6] band origins obtained from low- or medium-resolution spectra, calculated origins from Table 7 of [1], and also calculated origins obtained by simply summing the observed fundamentals (taken from Table 4 of [1]). Table 4 includes most (but not all) of our stronger bands. For the remaining 11 bands (mostly weak), Table 5 gives possible assignments which rely on the 'sums of fundamentals', since complete calculated model results from [1] are not available.

In Table 5, the A$_2$ symmetry state at 2651.502 cm$^{-1}$ is a special case since it was only observed through its perturbation of the B$_1$ 2651.032 cm$^{-1}$ state by means of a $c$-type Coriolis interaction (transitions to an A$_2$ vibration from the ground state are forbidden, as mentioned above). Rotational constants for the A$_2$ state were fixed at reasonable arbitrary values (see Table S2).

The relatively strong 3052.260 cm$^{-1}$ band in Table 5 is another special case. It must correspond to the feature reported at 3052.2 cm$^{-1}$ in [6], which was assigned there as $\nu_8 + \nu_9 + \nu_{11}$ (a B$_2$ mode) and later reassigned in [1] as $\nu_5 + \nu_8 + 2\nu_{12}$ (a B$_1$ mode). But our strong band is clearly due to an A$_1$ mode, so these previous assignments must be incorrect. Possible A$_1$ assignments for 3052.260 cm$^{-1}$ are shown in Table 5, but we have no good way to choose between them. Meanwhile, there is a much weaker B$_1$ band at 3050.961 cm$^{-1}$ which we assign as the real $\nu_5 + \nu_8 + 2\nu_{12}$ transition (see Table 4).

Three other reasonably strong bands (2240.061, 2595.265, 2651.032 cm$^{-1}$) do not match anything in the extensive Table 7 of [1], which includes about 94 DFE monomer bands between 437 and 3504 cm$^{-1}$. To explain 2240.061 cm$^{-1}$, we suggest two possible B$_1$ vibrations (see Table 5) and think that $\nu_4 +$ $v_6 + v_{12}$ is more likely. For 2595.265 cm$^{-1}$, we suggest three possible assignments of which $v_2 + 2v_{10}$ is perhaps most likely. And for the 2651.033 cm$^{-1}$ band, $v_3 + v_8$ seems to be very likely.

The fitted lower state rotational parameters of the three weak hot bands mentioned above (2253.0648, 2643.9648, 3096.6238 cm$^{-1}$) agree very well with precise microwave results [5] for the lowest excited vibrational state, $v_{10} = 1$ at 436.8851 cm$^{-1}$. These hot bands were $a$-type, which means that the excited states have B$_1$ symmetry, since $v_{10} = 1$ has B$_1$ symmetry. To assign the hot band upper vibrational states, we assumed that they were related to the nearby cold bands at 2250.273, 2648.113 and 3098.498 cm$^{-1}$, respectively (see Table 5). These connections were supported by the observed changes in rotational constants between lower and upper states ($\alpha$ values), which were similar for each of the cold and hot pairs.

## VI.    Conclusions

Low temperature ($\approx$ 2.3 K) supersonic jet spectra of 1,1-difluoroethylene (DFE) have been observed in the 2200 – 3100 cm$^{-1}$ region using a tunable optical parametric oscillator source to probe a pulsed slit jet. Spectra were observed of the dimer [DFE]$_2$ and the helium complexes DFE-He and DFE-He$_2$, all for the first time. In addition, a large number of bands in the region between 2200 and 3100 cm$^{-1}$ were observed for the DFE monomer. These results should be useful since DFE is a prototype for the study of vibrational resonances and anharmonicity in a "medium" size molecule [1,6].

DFE dimer has a slipped antiparallel structure with two-fold rotational symmetry around the $c$-inertial axis. Depending on the different levels of theory employed, our *ab initio* calculations indicated that the global minimum structure could have C$_{2h}$ symmetry or it could be slightly twisted so that the four C atoms are not co-planar (C$_2$ symmetry); in the latter case anyway the computed energy barrier at the more symmetric configuration is low enough that the effective zero-point average structure has C$_{2h}$ symmetry. Five bands of (DFE)$_2$ associated with different DFE monomer vibrations were observed with sufficient detail for rotational analysis, and they were red-shifted by from 3.5 to 6.7 cm$^{-1}$ relative to the monomer. The experimental value for the monomer center of mass separation of (DFE)$_2$ was 3.44 Å.

DFE-He has a structure with the He atom located at the F atom end, out of the DFE monomer plane, analogous to the previously observed DFE-Ne and -Ar complexes [23,24]. We observed a doubling of the lines in the spectra due to the effects of He atom tunneling between equivalent locations on either side of the monomer plane. Ten different bands of DFE-He were rotationally analyzed, and systematic residuals in these fits suggested that additional large amplitude tunneling or internal rotation motions are present. Vibrational frequency shifts relative to the monomer were very small (<0.06 cm$^{-1}$).

In DFE-He$_2$, the second He atom occupies the equivalent location on the other side of the monomer, so tunneling through the DFE plane is quenched. The lines of DFE-He$_2$ were relatively weak and only one band was analyzed. The experimental structure derived for DFE-He$_2$ located the He atoms slightly further from the monomer symmetry axis than was the case in DFE-He, which may be an effect of the absence of tunneling.

## Supporting Information

The Supporting Information contains equilibrium geometries (optimized at the rev-DSDPBEP86 and B2PLYP levels of theory) and their corresponding binding energies (computed using the jun-ChS extrapolation scheme) for the lowest energy isomers of the DFE dimers and molecular parameters for the observed bands of DFE monomer (see text for details). Also given are details of the fitted parameters for 23 bands of DFE monomer (Table S2), and for the 2219.66 cm$^{-1}$ band of DFE-He (Table S3).

## Acknowledgement

The financial support of the Natural Sciences and Engineering Research Council of Canada is gratefully acknowledged. One of the authors (ACP) gratefully acknowledges the CINECA award under the ISCRA initiative, for the availability of high-performance computing resources and support, and the financial support by ADir Funds. We thank R.A. Peebles for providing the DFE sample used in this work.

Low temperature jet spectra of (DFE)$_2$, DFE-He, DFE-He$_2$ and DFE in the 2210 – 3105 cm$^{-1}$ region (DFE = 1,1 difluoroethylene)

A. J. Barclay, A. R. W. McKellar, A. Pietropolli Charmet, and N. Moazzen-Ahmadi

## Table S1: Geometries for (CH2=CF2)2 complexes

1) the geometries of the dimers listed below (in Å) were optimised at revDSD-PBEP86-D3BJ/jun-cc-pVTZ level, using tight criteria and a SUPERFINE GRID.

2) the binding energies (not including ZPVE correction) were computed by CBS extrapolation.

| Structure I | | |
|---|---|---|
| F | -0.483702 | 2.009453 | -0.980099 |
| C | 0.319522 | 1.620525 | -0.007933 |
| C | 1.571609 | 1.231554 | -0.164245 |
| H | 2.002578 | 1.216146 | -1.151933 |
| H | 2.143558 | 0.927913 | 0.697272 |
| F | - 0.319522 | 1.679799 | 1.145402 |
| F | 0.319522 | -1.679799 | 1.145402 |
| C | -0.319522 | -1.620525 | -0.007933 |
| C | -1.571609 | -1.231554 | -0.164245 |
| H | -2.143558 | -0.927913 | 0.697272 |
| H | -2.002578 | -1.216146 | -1.151933 |
| F | 0.483702 | -2.009453 | -0.980099 |

Rotational constants (GHz):  A = 2.7755369   B = 1.1086994    C = 1.1067539

Binding energy : -2.25 kcal/mole        Electric dipole moment (Debye): 0.311

| Structure II | | |
|---|---|---|
| F | -1.518241 | 1.279188 | 0.148779 |
| C | -1.680791 | -0.028062 | 0.235880 |
| C | -1.331019 | -0.768654 | 1.271896 |
| H | -0.875606 | -0.299372 | 2.128463 |
| H | -1.503373 | -1.832172 | 1.245431 |
| F | -2.241234 | -0.479923 | -0.868464 |
| F | 2.125256 | -1.145256 | 0.230415 |
| C | 1.709640 | 0.046385 | -0.155212 |
| C | 1.978234 | 1.183032 | 0.460654 |
| H | 2.585388 | 1.172781 | 1.351090 |
| H | 1.584329 | 2.104992 | 0.065399 |
| F | 0.984539 | -0.069836 | -1.251807 |

Rotational constants (GHz): A = 2.9860105    B = 1.0766525    C = 1.0382416

Binding energy : -1.94  kcal/mole     Electric dipole moment (Debye):  1.726

| Structure III | | |
|---|---|---|
| F | -1.416242 | 0.255758 | 1.257375 |
| C | -1.964452 | 0.040183 | 0.077044 |
| C | -2.759857 | -0.969024 | -0.223665 |
| H | -2.998729 | -1.698661 | 0.532556 |
| H | -3.155196 | -1.053187 | -1.222609 |
| F | -1.598159 | 0.995315 | -0.760274 |
| F | 3.010333 | -0.735495 | 0.149759 |
| C | 1.922267 | -0.026264 | -0.086573 |
| C | 1.836315 | 1.288619 | -0.006991 |
| H | 2.707423 | 1.858802 | 0.271607 |
| H | 0.896586 | 1.771612 | -0.218449 |
| F | 0.931210 | -0.835540 | -0.415969 |

Rotational constants (GHz): A = 3.8293885   B = 0.8815837   C = 0.8277922

Binding energy : -1.73 kcal/mole        Electric dipole moment (Debye):  0.962

| Structure IV | | |
|---|---|---|
| F | 0.994493 | -0.519856 | -1.115030 |
| C | 1.734709 | 0.062724 | -0.190245 |
| C | 1.935424 | 1.364214 | -0.089466 |
| H | 1.472419 | 2.027122 | -0.801856 |
| H | 2.560384 | 1.744514 | 0.701939 |
| F | 2.241796 | -0.850502 | 0.613719 |
| F | -0.994492 | -0.519843 | 1.115039 |
| C | -1.734709 | 0.062721 | 0.190244 |
| C | -1.935439 | 1.364208 | 0.089453 |
| H | -1.472448 | 2.027128 | 0.801842 |
| H | -2.560400 | 1.744494 | -0.701958 |
| F | -2.241781 | -0.850517 | -0.613715 |

Rotational constants (GHz): A = 3.1386745   B = 1.0681243   C = 1.0020206

Binding energy : -1.67 kcal/mole        Electric dipole moment (Debye):  2.454

1) the geometry of the dimers listed below (in Å) were optimised at B2PLYP-D3BJ/jun-cc-pVTZ level, using tight criteria and a SUPERFINE GRID.

2) the binding energy (not including ZPVE correction) reported below only for geometry I was computed by CBS extrapolation.

| Structure I | | | |
|---|---|---|---|
| F | 0.359412 | 1.887933 | 1.079380 |
| C | -0.359397 | 1.633748 | 0.000001 |
| C | -1.601237 | 1.195019 | 0.000017 |
| H | -2.099906 | 1.014411 | 0.935080 |
| H | -2.099919 | 1.014382 | -0.935034 |
| F | 0.359397 | 1.887899 | -1.079396 |
| F | -0.359397 | -1.887899 | -1.079396 |
| C | 0.359397 | -1.633748 | 0.000001 |
| C | 1.601237 | -1.195019 | 0.000017 |
| H | 2.099919 | -1.014382 | -0.935034 |
| H | 2.099906 | -1.014411 | 0.935080 |
| F | -0.359412 | -1.887933 | 1.079380 |

Rotational constants (GHz): A = 2.7576869    B = 1.0855123    C = 1.0835002

Binding energy : -2.24 kcal/mole

| Structure II | | | |
|---|---|---|---|
| F | -1.534441 | 1.278867 | 0.121825 |
| C | -1.699089 | -0.027802 | 0.236110 |
| C | -1.348505 | -0.746778 | 1.282814 |
| H | -0.890264 | -0.264686 | 2.127153 |
| H | -1.521604 | -1.807746 | 1.279286 |
| F | -2.264499 | -0.499698 | -0.859308 |
| F | 2.171890 | -1.127701 | 0.252790 |
| C | 1.727994 | 0.048600 | -0.153911 |
| C | 1.974862 | 1.200156 | 0.435729 |
| H | 2.585944 | 1.222915 | 1.319941 |
| H | 1.559743 | 2.103718 | 0.027363 |
| F | 0.997562 | -0.106941 | -1.243995 |

Rotational constants (GHZ): A = 2.9987407    B = 1.0573125    C = 1.0194810

| Structure III | | | |
|---|---|---|---|
| F | -1.449050 | 0.181448 | 1.278758 |
| C | -1.975064 | 0.037052 | 0.076077 |
| C | -2.763185 | -0.948193 | -0.299612 |
| H | -3.018082 | -1.720362 | 0.403456 |
| H | -3.138443 | -0.972982 | -1.306610 |
| F | -1.590262 | 1.041505 | -0.695720 |
| F | 3.022279 | -0.730596 | 0.146289 |
| C | 1.931838 | -0.021655 | -0.089806 |
| C | 1.844528 | 1.290121 | -0.014629 |
| H | 2.712397 | 1.862530 | 0.259032 |
| H | 0.906624 | 1.771292 | -0.224352 |
| F | 0.940233 | -0.835070 | -0.414183 |

Rotational constants (GHz): A = 3.8213266    B = 0.8750023        C = 0.8209231

| **Structure IV** | | | |
|---|---|---|---|
| F | 1.034761 | 1.085528 | 0.518842 |
| C | 0.060311 | 1.761452 | -0.064517 |
| C | -0.060311 | 1.945810 | -1.363428 |
| H | 0.673669 | 1.526527 | -2.027370 |
| H | -0.889861 | 2.514070 | -1.743706 |
| F | -0.770102 | 2.222094 | 0.852130 |
| F | -1.034761 | -1.085528 | 0.518842 |
| C | -0.060311 | -1.761452 | -0.064517 |
| C | 0.060311 | -1.945810 | -1.363428 |
| H | -0.673669 | -1.526527 | -2.027370 |
| H | 0.889861 | -2.514070 | -1.743706 |
| F | 0.770102 | -2.222094 | 0.852130 |

Rotational constants (GHZ): A = 3.1537288    B = 1.0505406        C = 0.9849922

Table S2. Molecular parameters for the observed bands of DFE monomer (in cm$^{-1}$).[a]

| | $\sigma_0$ | $A$ | $B$ | $C$ | int | $n$ | rms |
|---|---|---|---|---|---|---|---|
| A$_1$ | ground | 0.367005884 | 0.347873582 | 0.178302611 | | | |
| A$_1$ | 2216.5815(2) | 0.366873(84) | 0.347257(51) | 0.177411(13) | 0.01 | 12 | 0.00027 |
| B$_1$ | 2219.6596(1) | 0.3651127(34) | 0.3464593(36) | 0.1769023(10) | 0.65 | 118 | 0.00020 |
| B$_1$ | 2240.0610(1) | 0.3656362(84) | 0.3470912(68) | 0.1778133(25) | 0.15 | 70 | 0.00018 |
| A$_1$ | 2246.4964(5) | 0.36711(24) | 0.34686(13) | 0.177494(31) | 0.01 | 10 | 0.00049 |
| A$_1$ | 2250.2726(1) | 0.366916(10) | 0.3471637(80) | 0.1774350(26) | 0.60 | 115 | 0.00041 |
| A$_1$ | 2592.9314(1) | 0.3655600(33) | 0.3460829(30) | 0.1766719(11) | 0.75 | 114 | 0.00016 |
| A$_1$ | 2595.2645(1) | 0.366768(10) | 0.3460239(90) | 0.1765378(29) | 0.09 | 60 | 0.00024 |
| A$_1$ | 2648.1128(1) | 0.3653232(95) | 0.3460409(82) | 0.1766851(21) | 1.2 | 157 | 0.00049 |
| B$_1$[b] | 2651.0319(1) | 0.366671(24) | 0.348993(38) | 0.178255(11) | 0.11 | 62 | 0.00091 |
| A$_2$[b] | 2651.5015(13) | [0.3662] | [0.3475] | [0.1778] | 0.0 | | |
| B$_1$ | 2673.0254(3) | 0.368672(59) | 0.345471(63) | 0.176585(20) | 0.02 | 27 | 0.00051 |
| B$_1$ | 2673.8630(2) | 0.3671745(96) | 0.3466188(95) | 0.1776428(26) | 0.3 | 106 | 0.00036 |
| A$_1$ | 2777.6991(1) | 0.366782(14) | 0.347195(14) | 0.1784553(58) | 0.08 | 53 | 0.00027 |
| B$_1$ | 3016.6176(1) | 0.3654825(62) | 0.3457391(65) | 0.1765435(16) | 0.09 | 82 | 0.00020 |
| A$_1$ | 3047.8403(3) | 0.36626 (22) | 0.34657(17) | 0.177958(28) | 0.012 | 16 | 0.00078 |
| B$_1$ | 3050.9606(2) | 0.36614(20) | 0.346279(98) | 0.177207(30) | 0.017 | 22 | 0.00044 |
| A$_1$ | 3052.2603(1) | 0.365626(17) | 0.346115(13) | 0.1785328(34) | 0.30 | 90 | 0.00040 |
| A$_1$ | 3058.1022(1) | 0.3664112(70) | 0.3469430(56) | 0.1781011(34) | 1.0 | 93 | 0.00026 |
| A$_1$ | 3098.4979(1) | 0.367224(26) | 0.347111(21) | 0.1777351(79) | 0.23 | 66 | 0.00056 |
| A$_1$ | 3101.3676(1) | 0.367039(22) | 0.346593(20) | 0.1782422(99) | 0.40 | 65 | 0.00059 |
| B$_1$[c] | 2689.9319(2) | 0.367116(23) | 0.346772(33) | 0.1770843(81) | | 25 | 0.00031 |
| B$_1$[c] | 3080.8499(1) | 0.365648(37) | 0.345770(29) | 0.1761832(99) | | 42 | 0.00048 |
| B$_1$[c] | 3533.5084(2) | 0.367464(43) | 0.346787(39) | 0.177000(23) | | 22 | 0.00036 |

[a] The ground state parameters are from [25]. int is the approximate relative intensity, $n$ is the number of fitted transitions, and rms is the root mean square deviation of the fit.

[b] The A$_2$ state at 2651.5015 cm$^{-1}$ is observed only via its perturbation of 2651.0319 cm$^{-1}$ band. These states are connected by a $c$-type Coriolis interaction with matrix element equal to ½√[$J(J+1) - K(K+1)$] × 0.01730(15) cm$^{-1}$.

[c] These are *a*-type hot bands originating in the $\nu_{10} = 1$ state, which has $B_1$ symmetry and is located at 436.8851 cm$^{-1}$. The band origins are 2253.0468(2), 2643.9648(1), and 3096.6238(2) cm$^{-1}$.

Table S3. Observed transitions in the 2219.6 cm$^{-1}$ band of DFE-He (in cm$^{-1}$).

|  | J´ | Ka´ | Kc´ | J" | Ka" | Kc" | Observed | Calculated | Obs-calc |
|---|---|---|---|---|---|---|---|---|---|
| (+) | 0 | 0 | 0 | 1 | 1 | 1 | 2219.2673 | 2219.2738 | -0.0065 |
| (+) | 1 | 0 | 1 | 1 | 1 | 0 | 2219.5789 | 2219.5854 | -0.0065 |
| (+) | 1 | 0 | 1 | 2 | 1 | 2 | 2218.9606 | 2218.9606 | 0.0000 |
| (+) | 1 | 1 | 0 | 1 | 0 | 1 | 2219.7314 | 2219.7261 | 0.0054 |
| (+) | 1 | 1 | 0 | 2 | 2 | 1 | 2218.8216 | 2218.8199 | 0.0017 |
| (+) | 1 | 1 | 1 | 0 | 0 | 0 | 2220.0445 | 2220.0378 | 0.0067 |
| (+) | 1 | 1 | 1 | 2 | 0 | 2 | 2219.0378 | 2219.0377 | 0.0001 |
| (+) | 1 | 1 | 1 | 2 | 2 | 0 | 2218.7919 | 2218.7896 | 0.0022 |
| (+) | 2 | 0 | 2 | 1 | 1 | 1 | 2220.2693 | 2220.2699 | -0.0006 |
| (+) | 2 | 0 | 2 | 2 | 1 | 1 | 2219.5525 | 2219.5537 | -0.0012 |
| (+) | 2 | 0 | 2 | 3 | 1 | 3 | 2218.6556 | 2218.6544 | 0.0012 |
| (+) | 2 | 1 | 1 | 2 | 0 | 2 | 2219.7527 | 2219.7535 | -0.0008 |
| (+) | 2 | 1 | 1 | 2 | 2 | 0 | 2219.5071 | 2219.5054 | 0.0017 |
| (+) | 2 | 1 | 1 | 3 | 2 | 2 | 2218.5084 | 2218.5085 | -0.0001 |
| (+) | 2 | 1 | 2 | 1 | 0 | 1 | 2220.3485 | 2220.3467 | 0.0018 |
| (+) | 2 | 1 | 2 | 3 | 0 | 3 | 2218.6978 | 2218.6990 | -0.0011 |
| (+) | 2 | 2 | 0 | 1 | 1 | 1 | 2220.5136 | 2220.5138 | -0.0003 |
| (+) | 2 | 2 | 1 | 1 | 1 | 0 | 2220.4842 | 2220.4838 | 0.0004 |
| (+) | 3 | 0 | 3 | 2 | 1 | 2 | 2220.6011 | 2220.5999 | 0.0013 |
| (+) | 3 | 0 | 3 | 3 | 1 | 2 | 2219.4975 | 2219.4962 | 0.0013 |
| (+) | 3 | 1 | 3 | 2 | 0 | 2 | 2220.6435 | 2220.6447 | -0.0012 |
| (+) | 3 | 2 | 2 | 2 | 1 | 1 | 2220.7912 | 2220.7898 | 0.0014 |
| (+) | 4 | 1 | 3 | 4 | 2 | 2 | 2219.4952 | 2219.4946 | 0.0006 |
| (+) | 4 | 2 | 3 | 3 | 1 | 2 | 2221.0785 | 2221.0812 | -0.0026 |
| (+) | 4 | 4 | 0 | 5 | 5 | 1 | 2217.4598 | 2217.4602 | -0.0004 |
| (+) | 4 | 4 | 1 | 3 | 3 | 0 | 2221.3464 | 2221.3493 | -0.0029 |
| (+) | 4 | 4 | 1 | 5 | 5 | 0 | 2217.4598 | 2217.4599 | -0.0001 |
| (+) | 5 | 5 | 0 | 4 | 4 | 1 | 2221.7476 | 2221.7469 | 0.0006 |
| (+) | 5 | 5 | 1 | 4 | 4 | 0 | 2221.7476 | 2221.7467 | 0.0009 |
| (-) | 0 | 0 | 0 | 1 | 1 | 1 | 2219.2830 | 2219.2782 | 0.0048 |
| (-) | 1 | 0 | 1 | 1 | 1 | 0 | 2219.5970 | 2219.5906 | 0.0064 |
| (-) | 1 | 0 | 1 | 2 | 1 | 2 | 2218.9606 | 2218.9627 | -0.0021 |
| (-) | 1 | 1 | 0 | 1 | 0 | 1 | 2219.7144 | 2219.7201 | -0.0058 |
| (-) | 1 | 1 | 0 | 2 | 2 | 1 | 2218.8315 | 2218.8342 | -0.0028 |
| (-) | 1 | 1 | 1 | 0 | 0 | 0 | 2220.0302 | 2220.0342 | -0.0040 |
| (-) | 1 | 1 | 1 | 2 | 2 | 0 | 2218.8040 | 2218.8067 | -0.0027 |
| (-) | 2 | 0 | 2 | 1 | 1 | 1 | 2220.2760 | 2220.2735 | 0.0025 |
| (-) | 2 | 0 | 2 | 2 | 1 | 1 | 2219.5627 | 2219.5598 | 0.0028 |
| (-) | 2 | 0 | 2 | 3 | 1 | 3 | 2218.6512 | 2218.6537 | -0.0025 |
| (-) | 2 | 1 | 1 | 2 | 0 | 2 | 2219.7435 | 2219.7440 | -0.0005 |
| (-) | 2 | 1 | 1 | 3 | 2 | 2 | 2218.5195 | 2218.5173 | 0.0022 |
| (-) | 2 | 1 | 2 | 3 | 0 | 3 | 2218.6978 | 2218.6958 | 0.0019 |
| (-) | 2 | 2 | 0 | 1 | 1 | 1 | 2220.5029 | 2220.5012 | 0.0017 |
| (-) | 2 | 2 | 0 | 2 | 1 | 1 | 2219.7900 | 2219.7875 | 0.0025 |
| (-) | 2 | 2 | 1 | 1 | 1 | 0 | 2220.4751 | 2220.4736 | 0.0015 |
| (-) | 3 | 0 | 3 | 2 | 1 | 2 | 2220.6011 | 2220.6035 | -0.0024 |

| (-) | 3 | 1 | 2 | 3 | 0 | 3 | 2219.7900 | 2219.7881 | 0.0020 |
|-----|---|---|---|---|---|---|-----------|-----------|---------|
| (-) | 3 | 1 | 3 | 2 | 0 | 2 | 2220.6490 | 2220.6453 | 0.0037 |
| (-) | 3 | 2 | 2 | 2 | 1 | 1 | 2220.7798 | 2220.7803 | -0.0005 |
| (-) | 4 | 2 | 3 | 3 | 1 | 2 | 2221.0695 | 2221.0736 | -0.0041 |
| (-) | 4 | 4 | 0 | 3 | 3 | 1 | 2221.3392 | 2221.3401 | -0.0010 |
| (-) | 4 | 4 | 0 | 5 | 5 | 1 | 2217.4774 | 2217.4776 | -0.0001 |
| (-) | 4 | 4 | 1 | 3 | 3 | 0 | 2221.3392 | 2221.3387 | 0.0004 |
| (-) | 4 | 4 | 1 | 5 | 5 | 0 | 2217.4774 | 2217.4773 | 0.0001 |
| (-) | 5 | 5 | 0 | 4 | 4 | 1 | 2221.7419 | 2221.7418 | 0.0001 |
| (-) | 5 | 5 | 1 | 4 | 4 | 0 | 2221.7419 | 2221.7416 | 0.0004 |